\documentclass[showpacs, nofootinbib, preprintnumbers,amsmath,amssymb, prd]{revtex4}
\usepackage{amsmath,amssymb,graphicx}
\usepackage{epsf}
\usepackage{epstopdf}
\usepackage {amssymb}
\newcommand{\nc}{\newcommand}
\nc{\ba}{\begin{eqnarray}}
\nc{\ea}{\end{eqnarray}}
\newcommand\be{\begin{equation}}
\newcommand\ee{\end{equation}}

\newcommand\mPl{{m_{\rm Pl}}}
\newcommand\fNL{{f_{\rm NL}}}
\newcommand\F{{\cal{F}}}

\nc{\x}{{\bf{x}}}

\begin{document}


\title{Curvaton and the inhomogeneous end of inflation}

\author{Hooshyar Assadullahi$^{1, 2}$}
\email{hooshyar.assadullahi-AT-port.ac.uk}
\author{Hassan Firouzjahi$^{3}$}
\email{firouz-AT-mail.ipm.ir}
\author{Mohammad Hossein Namjoo$^{4, 5}$}
\email{mh.namjoo-AT-mail.ipm.ir}
\author{David Wands$^{1}$}
\email{david.wands-AT-port.ac.uk}
\affiliation{$^1$ Institute of Cosmology and Gravitation, University of Portsmouth, Dennis Sciama Building, Burnaby Road, Portsmouth PO1 3FX, United Kingdom}
\affiliation{$^2$ School of Earth and Environmental Sciences, University of Portsmouth, Burnaby Building, Burnaby Road, Portsmouth PO1 3QL, United Kingdom}
\affiliation{$^3$School of Astronomy, Institute for Research in
Fundamental Sciences (IPM),
P. O. Box 19395-5531,
Tehran, Iran}
\affiliation{$^4$Yukawa Institute for theoretical Physics,
 Kyoto University, Kyoto 606-8502, Japan}
\affiliation{$^5$School of Physics, Institute for Research in
Fundamental Sciences (IPM),
P. O. Box 19395-5531,
Tehran, Iran}


\begin{abstract}
\vspace{0.3cm}
We study the primordial density perturbations and non-Gaussianities generated from the combined effects of an inhomogeneous end of inflation and curvaton decay in hybrid inflation. This dual role is played by a single isocurvature field which is massless during inflation but acquire a mass at the end of inflation via the waterfall phase transition. We calculate the resulting primordial non-Gaussianity characterized by the non-linearity parameter, $\fNL$, recovering the usual end-of-inflation result when the field decays promptly and the usual curvaton result if the field decays sufficiently late.

\vspace{0.3cm}

\end{abstract}

\date\today

\preprint{ IPM/A-2012/013, YITP-12-65 }

\maketitle

\section{Introduction}

There are many mechanisms by which quantum fluctuations of light fields present during inflation could affect the primordial density perturbation \cite{Bassett:2005xm, Lyth-Liddle}. Variations in the local value of an inflaton field, slow-rolling during inflation, lead to variations in the local duration of inflation and hence the resulting local density after inflation has ended. But local variations in other fields, not necessarily evolving during inflation, could also affect the local expansion either during or after inflation. Variations orthogonal to the background field evolution during inflation have been characterized as entropy field perturbations \cite{Gordon:2000hv} which can alter gauge-invariant density perturbations on very large (super-Hubble) scales even after inflation has ended due to variations in the local equation of state and non-adiabatic pressure perturbations \cite{Wands:2000dp}.
While adiabatic field fluctuations in a single canonical inflaton field lead to Gaussian, adiabatic density perturbations after inflation, entropy field fluctuations during inflation may lead to a much richer phenomenology of non-Gaussian and non-adiabatic primordial density perturbations \cite{Salopek:1988qh,Linde:1996gt,Langlois:1999dw,Lyth:2002my,Bartolo:2003jx,Bassett:2005xm}.

A simple example of how isocurvature field perturbations can affect the primordial density perturbation is by altering the time at which inflation ends in hybrid inflation models where the end of inflation is triggered by an tachyonic instability in a ``waterfall'' field, leading to a rapid phase transition from false to true vacuum state \cite{Linde:1991km,Linde:1993cn,Copeland:1994vg}. If the local variations of an isocurvature field, coupled to the waterfall field, alter the point at which the false vacuum becomes unstable then they will lead to a primordial density perturbation~\cite{Bernardeau:2004zz,Lyth:2005qk,Dvali:2003em,Sasaki:2008uc, Naruko:2008sq, Huang:2009vk, Yokoyama:2008xw, Emami:2011yi, Lyth:2012br}. On the other hand, in the curvaton scenario~\cite{Lyth:2001nq,Moroi:2001ct} the isocurvature field is only weakly-coupled to the inflaton and its decay products, so that it survives after inflation has ended and may source the primordial density perturbation when it decays into radiation some time after inflation has ended.

In this paper we will study the effect of entropy fluctuations in a massless field during hybrid inflation, in a model originally considered by Lyth~\cite{Lyth:2005qk}. We will show how primordial density perturbations can be generated from inflaton field fluctuations and entropy fluctuations in the isocurvature field which is coupled to the waterfall field. Entropy fluctuations therefore affect the surface of end of inflation,
{\em and} the field spontaneously acquires a mass at the waterfall transition. Therefore this massless field during inflation can play the role of a curvaton field which can have a significant energy density when the field oscillations decay some time after inflation.

We calculate the primordial density perturbations from all these effects using the $\delta N$ formalism \cite{Starobinsky:1986fxa,Sasaki:1995aw, Wands:2000dp, Lyth:2005fi}. We identify the non-linear dimensionless density perturbation, $\zeta$ \cite{Bardeen:1983qw,Bardeen:1988hy,Malik:2008im}, with the perturbation in the local integrated expansion, $\delta N$, from an initial uniform-curvature hypersurface up to a uniform-density hypersurface \cite{Lyth:2004gb} in the long-wavelength limit~\cite{Salopek:1990jq} where the local expansion is a function of the local field fluctuations, $\delta\phi_i$, on the initial hypersurface:
\ba
\label{p-q}
\zeta  =
 \sum_i N_{,i} \delta\phi_i + \frac12 \sum_{i,j} N_{,ij} \delta\phi_i \delta\phi_j + \ldots \,.
\ea
where $N_{,i}\equiv \partial N/\partial\phi_i$.
We consider the case of canonical light scalar fields, $\phi_i$, where all these field fluctuations have a Gaussian distribution with power spectrum
${\cal{P}}_{\delta \phi_i} \simeq (\frac{H_k}{2\pi})^2$ when the wave-mode of interest, $k$, leaves the Hubble-horizon, $k=aH$, with the Hubble expansion rate $H_k$.
The primordial power spectrum is then given at leading order by
\ba
\label{power0}
{\cal{P}}_{\zeta} = \left(\frac{H_k}{2\pi} \right)^2 \sum_i N_{,i}^2
\ea
and the lowest-order non-Gaussianity parameter, $f_{NL}$, is given by \cite{Lyth:2005fi}
\ba
\label{fNL-def}
\frac{6}{5} f_{NL} = \frac{\sum_{,ij} N_{,i}N_{,j} N_{,ij}}{\left(\sum_j N_{,j}^2 \right)^2} \, .
\ea

We introduce the hybrid inflation model in Section~II and discuss the inflaton field dynamics and perturbations. In Section~III we discuss the end of inflation and the density perturbations produced due to the inhomogeneous end of inflation. In Section IV we discuss how the same isocurvature field can act as a curvaton field, producing additional density perturbations when it finally decays some time after inflation. We present our results for observable quantities, such as the tilt of the primordial power spectrum and the non-linearity parameter, $\fNL$, in Section~V. We conclude in Section~VI.

\section{Hybrid Inflation}

Our specific hybrid inflation model contains three fields, the inflaton field $\phi$, the waterfall field $\chi$ and the light isocurvature scalar field $\sigma$, which we will refer to as the curvaton field. The potential is
\ba
\label{potential}
V= \frac{\lambda}{4} \left( \chi^2 - \frac{M^2}{\lambda} \right)^2  + \frac{m^2}{2} \phi^2
+ \frac{g^2}{2} \phi^2 \chi^2 + \frac{\gamma^2}{2} \chi^2 \sigma^2 \, ,
\ea
where $\gamma^2, g^2$ and $\lambda$ are dimensionless couplings. This is the same as the original hybrid inflation model considered in Refs.~\cite{Linde:1993cn,Copeland:1994vg} except that we have included the additional field, $\sigma$, coupled to the waterfall field, as proposed by Lyth \cite{Lyth:2005qk}. The is the same as the model considered by Lyth, except that we have assumed for simplicity that there is no additional self-interaction potential for the $\sigma$-field.

We assume that $\phi$ is initially large and the $\chi$ field rapidly rolls to its local minimum $\chi=0$ for $g^2\phi^2\gg M^2$. Inflation proceeds as the inflaton field, $\phi$, rolls towards $\phi=0$. Assuming the inflaton is light during inflation, $m^2\ll H^2$, we have slow-roll inflation. Without loss of generality we assume that $\phi>0$ during inflation.
Since $\sigma$ is effectively massless when $\chi^2=0$, we assume it remains fixed during inflation at some initial value $\sigma_*$.

We will introduce the dimensionless mass parameters $\alpha$ and $\beta$ for the inflaton and waterfall fields
\ba
\label{alpha-beta}
\alpha \equiv \dfrac{m^2}{H^2}=\dfrac{12 \lambda m^2 M_P^2}{ M^4}
\qquad , \qquad   \beta \equiv \dfrac{M^2}{H^2}=\dfrac{12 \lambda  M_P^2}{ M^2}\, ,
\ea
where $M_P = 1/\sqrt{8\pi G}$ is the reduced Planck mass.
The condition that the waterfall is heavy during inflation requires that $\beta\gg1$.
To sustain a period of slow-roll inflation we require that $\alpha \ll1$.
Furthermore, the condition that the waterfall phase transition at the end of inflation is very sudden requires that \cite{Abolhasani:2010kr, Abolhasani:2011yp}
$\alpha \beta >1$.

\subsection{Inflaton dynamics}

For simplicity, we consider the limit where the potential is dominated by the vacuum term so that
the Hubble parameter during inflation is approximately constant and given by
\begin{eqnarray}
H^2\simeq\frac{V(\phi,0)}{3M_{P}^2}
\simeq\frac{M^4}{12\lambda M_{P}^2}\, .
\label{hubble}
\end{eqnarray}

The attractor solution for the inflaton, $\phi$, in the vacuum-dominated limit is then
\ba
\label{phi-N}
\phi(N) = \phi_e e^{-r N} \, ,
\ea
where $N$ is the number of e-folds (the logarithmic expansion)
\ba
 \label{Nint}
N = \int_t^{t_e} H dt = \int_\phi^{\phi_e} \frac{H}{\dot\phi} d\phi \,,
\ea
and $\phi_e$ is the value of the inflaton field at the end of inflation.

In the slow-roll limit, $\alpha\ll1$, we have $r\simeq \alpha/3$ in Eq.~(\ref{phi-N}) and hence
\ba
\label{epsilon}
\epsilon \equiv -\frac{\dot{H}}{H^2}
 \simeq \frac{\alpha^2\phi^2}{18M_P^2} \,.
\ea
In order to satisfy the vacuum domination condition (\ref{hubble}), we require the ratio of inflaton energy density to the total potential, to be small, i.e., $m^2\phi^2/2\ll M^4/4\lambda$ which corresponds to $\epsilon\ll\alpha/3$.

\subsection{Inflaton perturbations during inflation}

During inflation only the inflaton field $\phi$ is classically evolving towards its final value. Therefore quantum fluctuations in the inflaton field give rise to adiabatic curvature perturbations on super-Hubble scales. Since the dimensionless density perturbation, $\zeta=\delta N$, remains constant for adiabatic perturbations on super-Hubble scales, it can be calculated in terms of the field values at Hubble-exit, $N_{,\phi}\delta\phi=N_{,\phi_*}\delta\phi_*=-H\delta\phi_*/\dot\phi_*$, etc, and thus
\be
\zeta_* = \delta N \simeq \dfrac{3}{\alpha\phi_\ast} \delta\phi_\ast - \dfrac{3}{2\alpha \phi_\ast^2} \delta\phi_\ast^2 + \ldots
\ee
In the slow-roll limit non-linear terms are negligible, $|N_{,\phi_\ast \phi_\ast}|\sim \alpha N_{,\phi_\ast}^2 \ll N_{,\phi_\ast}^2$, and henceforth we will treat the density perturbation due to inflaton fluctuations as an effectively Gaussian distribution, $\zeta_*\simeq(3/\alpha)\delta\phi_*/\phi_*$.

\section{End of inflation}

\subsection{Waterfall transition}

From Eq.~(\ref{potential}) the derivative of the potential with respect to the waterfall field
vanishes at $\chi=0$ and also at $\lambda\chi^2=M^2 - \gamma^2\sigma^2-g^2\phi^2$ for $g^2\phi^2<M^2 - \gamma^2\sigma^2$.
The effective mass of the $\chi$ field is given by
\be
V_{,\chi\chi} =
g^2\phi^2 + \gamma^2\sigma^2 - M^2 + 3\lambda \chi^2 \,.
\ee
Thus the $\chi$ field remains fixed at $\chi=0$ for $g^2\phi^2 > M^2 - \gamma^2\sigma^2$.

In the original model of hybrid inflation, where $\gamma=0$ \cite{Linde:1993cn, Copeland:1994vg}, inflation ends after the inflaton field reaches the critical value $\phi=\phi_c$, where
\be
\phi_c \equiv \frac{M}{g} \,.
\ee
For $\phi<\phi_c$ the waterfall field becomes tachyonic and it rolls to its global minima where $\phi=0$ and $|\chi| = M/\sqrt{\lambda}$. We require $\alpha\beta>1$ so that this transition is very sharp and inflation ends abruptly after the waterfall transition.

In the present model, due to the coupling of the waterfall field to the light field $\sigma$, the onset of the waterfall transition and hence the end of inflation is determined by the condition $\phi=\phi_e$ where
\ba
 \label{end-surface}
 \phi_e^2 + \frac{\gamma^2}{g^2} \sigma^2 = \phi_c^2 \, ,
\ea
and $\phi_e\leq \phi_c$ represents the value of inflaton field at the end of inflation which is now a function of the local value of the isocurvature field, $\sigma$.
It will be convenient to re-write this in a dimensionless form
\be
 \frac{\phi_e^2}{\phi_c^2} = 1 - \F \,,
 \ee
where
\be
 \label{defF}
 {\cal F} \equiv \dfrac{\gamma^2 \sigma_*^2}{g^2 \phi_c^2} \, ,
\ee
is a dimensionless measure of the effect of the curvaton field at the end of inflation. Note that $0\leq \F<1$\footnote{For $\gamma^2\sigma_*^2>g^2\phi_c^2$ (corresponding to $\F>1$) the waterfall field would remain fixed at $\chi=0$ even as $\phi\to0$ and the field $\sigma$ would then act as an inflaton field, slow-rolling towards its minimum to destabilize the waterfall field and end inflation.}. When $\F=0$ we recover the familiar hybrid inflation results where first-order density perturbations are solely due to inflaton fluctuations.

We will assume that the curvaton field, $\sigma$, is weakly coupled to the other fields so that its value remains fixed at the end of inflation when $\phi$ and $\chi$ evolve rapidly to settle at $\phi=0$ and
\be
 \langle \chi^2 \rangle = (1-\F) \frac{M^2}{\lambda} \,.
\ee
By assuming that the curvaton field evolves much more slowly than the inflaton and waterfall fields, we require that the curvaton effective mass remains much smaller than the effective masses of the inflaton and waterfall fields at the end of inflation and hence $\gamma^2\ll {\rm min}\{ g^2, \lambda \}$.

One important question in hybrid inflation is how inflation ends after the waterfall phase transition. The waterfall phase transition leads to two important back-reaction effects \cite{Abolhasani:2010kr, Abolhasani:2011yp}. Due to the interactions, $g^2\phi^2\chi^2$ and $\lambda\phi^4/4$, the tachyonic growth of the waterfall field $\chi$ can back-react on $\phi$ and on itself.
One natural criteria to determine end of inflation is when the coupling $g^2 \chi^2 \phi^2$
induces an effective mass for the inflaton larger then $H$ so that the inflaton field becomes too heavy to sustain slow-roll inflation. Comparing this induced mass to $H$ we have
\ba
\label{phic-Mp}
\frac{g^2 \langle \chi^2 \rangle }{H^2} = 12 (1-\F) g^2 \frac{M_P^2}{M^2} =
12 (1-\F) \left( \frac{M_P}{\phi_c} \right)^2 \, .
\ea
Therefore, in order for inflation to quickly end after the waterfall, one requires that \cite{Abolhasani:2011yp} $\phi_c\ll\sqrt{12(1-\F)} \, M_P$.

Given the potential (\ref{potential}), where $\chi$ is classically zero during inflation, the $\sigma$ field is massless during inflation. It acquires an effective mass, $m_\sigma$, spontaneously at the end of inflation due to the waterfall phase transition
\ba
\label{m-sigma}
m_\sigma^2 = \gamma^2  \langle \chi^2 \rangle = \frac{\gamma^2}{\lambda} (1-\F) M^2 \, .
\ea
To make sure that $\sigma$ does not trigger a second stage of inflation with the chaotic inflation potential $m_\sigma^2 \sigma^2/2$, we require that $\sigma$ is sub-Planckian at the end of inflation
\ba
\sigma_* < M_P \, .
\ea

Note that the curvaton effective mass is a function of $\langle \chi^2 \rangle$ and hence $\sigma$. Therefore the curvaton mass can change as the curvaton itself evolves. However this effect is only important for $\F\simeq1$ and the curvaton mass is effectively constant for $\F\ll1$.

Although the curvaton field is effectively massless and therefore has negligible energy density during inflation, $\rho_{\sigma\ast}\simeq0$, if the curvaton field suddenly acquires an effective mass (\ref{m-sigma}) at the end of inflation then it also acquires an energy density
\be
\rho_{\sigma e} = \frac12 m_\sigma^2\sigma_\ast^2 = \frac{\gamma^2}{2\lambda} (1-\F) M^2 \sigma_\ast^2 \, .
\ee
Therefore, the energy density suddenly acquired by the $\sigma$ field relative to the total energy density at the end of inflation is given by
\ba
\label{fL}
\Omega_{\sigma e}
\equiv \frac{\rho_{\sigma e}}{3\mPl^2H^2}
= \frac{2 \lambda \gamma^2 \langle \chi^2 \rangle \sigma_*^2}{M^4}
= 2\F (1-\F) \, .
\ea
Thus the fractional energy of the curvaton at the end of inflation is given by $\Omega_{\sigma e}\simeq2\F$ for $\F\ll1$. The fractional energy of the curvaton reaches a maximum, $\Omega_{\sigma e}=1/2$ for $\F=1/2$ and then drops to zero as $\F\to1$.

\subsection{Inhomogeneous end of inflation}

Local fluctuations in the curvaton field at the end of inflation, $\delta \sigma$, can induce a change in the integrated expansion, $\delta N$ in Eq.(\ref{p-q}), which results in an inhomogeneous curvature perturbation at the end of inflation, $\zeta_e$ \cite{Lyth:2005qk}.
{}From Eq.~(\ref{end-surface}), fluctuations in the field $\sigma$ lead to a change in the value of the inflaton field at the end of inflation,
\ba
\partial_\sigma \phi_e = - \frac{\gamma^2\sigma}{g^2\phi_e} \,.
\ea
This leads to fluctuations in the integrated expansion at the end of inflation. From Eq.~(\ref{Nint}) we have
\ba
N_{,\phi_e} = \frac{H}{\dot\phi_e} = \frac{-3}{\alpha\phi_e} \,.
\ea

Including both inflaton fluctuations during inflation and the fluctuations in the final value of the inflaton field at the end of inflation we obtain
\ba
\label{zeta-e}
\zeta_e = \delta N =  N_{, \phi_\ast} \delta \phi_\ast +
N_{, \phi_e} \partial_\sigma \phi_e \delta \sigma + \frac{1}{2} \left[  N_{,\phi_e \phi_e} {(\partial_\sigma \phi_e)}^2+N_{,\phi_e}\partial_\sigma^2 \phi_e
\right] \delta \sigma^2 + \ldots
\ea

\begin{figure}
\centering
\includegraphics{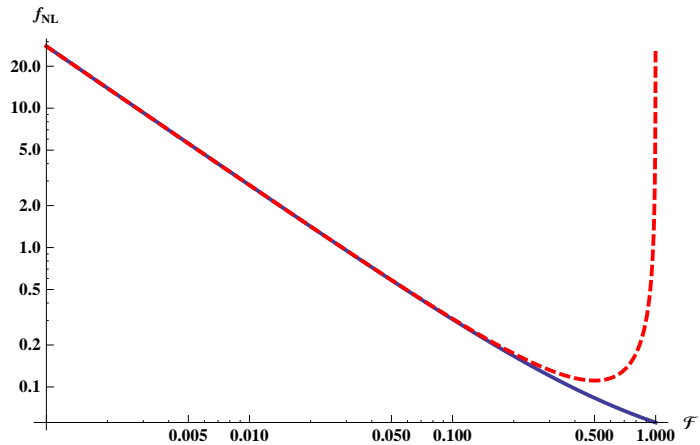}
\caption{The non-linearity parameter, $\fNL$ defined in Eq.~(\ref{fNL-def}), from the inhomogeneous end of inflation as a function of the dimensionless parameter, $\F$ defined in Eq.~(\ref{defF}), neglecting inflaton perturbations ($w_\sigma=1$). The exact result (\ref{final-zeta-e}) is shown by the solid blue line, while the approximate form (\ref{fNL-end}) is shown by the dashed red line.}
\label{End}
\vspace{0.5cm}
\end{figure}

Including all the relevant terms in Eq.~(\ref{zeta-e}) we then have \cite{Lyth:2005qk}
\ba
 \label{final-zeta-e}
\zeta_e
 &=& \dfrac{3\delta\phi_\ast}{\alpha\phi_\ast} + \dfrac{3}{\alpha } \dfrac{\gamma^2 \sigma}{g^2 \phi_e^2}\delta\sigma +
  \dfrac{3}{2 \alpha} \dfrac{\gamma^2}{g^2 \phi_e^2}
  \left( 1+2 \dfrac{\gamma^2 \sigma^2}{g^2 \phi_e^2}  \right) \delta\sigma^2 +\ldots
  \nonumber \,,\\
 &=& \dfrac{3}{\alpha} \left[ \frac{\delta\phi_\ast}{\phi_\ast} + \frac{\F}{1-\F} \frac{\delta\sigma}{\sigma}
  + \frac{{\F} (1+\F)}{2(1-\F)^2} \left( \frac{\delta\sigma}{\sigma} \right)^2 \right] +\ldots \,.
\ea
In figure~\ref{End} we illustrate the corresponding non-Gaussianity parameter (\ref{fNL-def}) as a function of $\F$.

\section{Curvaton decay}

The curvaton field, $\sigma$, is assumed to be weakly coupled and therefore, in contrast to the inflaton and waterfall fields, the curvaton decay time may be much longer than the Hubble time at the end of inflation, $\Gamma\ll H_*$. We may have an extended period of time after inflation, before the curvaton decays, and thus even if the curvaton density is initially negligible at the end of inflation ($\F(1-\F)\ll1$) it may become significant or even dominate the total energy density before it decays.

The non-linearly conserved radiation and matter perturbations, when the curvaton oscillates after the end of inflation but before the curvaton decays, can be given relative to the total curvature perturbation, $\zeta$, and written in terms of the radiation and matter density perturbations on the uniform-density hypersurface as~\cite{Lyth:2004gb}
 \ba
  \label{def-zetagamma}
  \zeta_\gamma &=& \zeta + \frac14 \ln \left( 1+ \frac{\delta\rho_\gamma}{\rho_\gamma} \right) \,, \\
 \label{def-zetasigma}
  \zeta_\sigma &=& \zeta + \frac13 \ln \left( 1+ \frac{\delta\rho_\sigma}{\rho_\sigma} \right) \,.
 \ea
In particular we have $\delta\rho_\gamma=-\delta\rho_\sigma$ on the uniform-density hypersurface at the end of inflation, and hence
\ba
 \label{zeta-gamma}
  \zeta_\gamma
 &=& \zeta_e + \frac14 \ln \left( 1 - \frac{\Omega_\sigma}{1-\Omega_\sigma} \frac{\delta\rho_\sigma}{\rho_\sigma} \right)
 \nonumber \,,
\\
   &=& \zeta_e - \frac{\F(1-\F)}{1-2\F(1-\F)} \frac{\delta\sigma}{\sigma} - \frac{\F(1-\F)(1+2\F(1-\F))}{2(1-2\F(1-\F))^2} \left( \frac{\delta\sigma}{\sigma} \right)^2 + \ldots
   \,.
\ea
where $\delta\sigma$ is the (isocurvature) curvaton field perturbation on uniform-density hypersurfaces.
Comparing this expression with Eq.~(\ref{final-zeta-e}) we see that $\zeta_\gamma-\zeta_e={\cal O}(\alpha)\zeta_e$. This difference is small in the slow-roll limit and is usually neglected. However in our case we wish to consistently allow for the effect of the $\sigma$ field both on the end of inflation and when it decays. Therefore we shall take account of this difference between the radiation density perturbation and the total density perturbation at the end of inflation, $\zeta_\gamma$ and $\zeta_e$, for non-zero $\F$.

In the sudden-decay approximation we can then derive the total non-linear perturbation, $\zeta$, in terms of $\zeta_\gamma$ and $\zeta_\sigma$ at the decay hypersurface corresponding to $H=\Gamma$ \cite{Sasaki:2006kq, Assadullahi:2007uw,  Langlois:2008vk}
\ba
\Omega_\gamma e^{4(\zeta_\gamma-\zeta)}+\Omega_{\sigma}e^{3(\zeta_\sigma-\zeta)} = 1 \,,
\ea
$\zeta$ is non-linearly conserved on large scales for adiabatic density perturbations after the curvaton has decayed.

At zero-order this requires simply $\Omega_\gamma+\Omega_\sigma=1$, while expanding up to second-order we have the non-linear perturbation
\ba
 \label{curvaton-zeta}
\zeta = \zeta_\gamma + f_d (\zeta_\sigma-\zeta_\gamma) + \frac{f_d(1-f_d)(3+f_d)}{2} (\zeta_\sigma-\zeta_\gamma)^2 + \ldots \,,
\ea
where
\ba
 \label{f-def}
f_d \equiv \left( \frac{3\Omega_\sigma}{4-\Omega_\sigma} \right)_{H=\Gamma} \,.
\ea

We will distinguish two cases according to whether the curvaton becomes underdamped and oscillates about its minimum immediately at the end of inflation ($m_\sigma^2>H_*^2$) or whether it remains overdamped at the end of inflation ($m_\sigma^2<H_*^2$) and only begins to oscillate when the Hubble rate drops below the curvaton mass some time after inflation. In either case we assume that the curvaton decay only happens once the field is oscillating ($\Gamma<m_\sigma$).

\subsection{Heavy curvaton after the end of inflation}

First let us assume that the curvaton effective mass is larger than the Hubble scale at the end of inflation, $m_\sigma>H_*$, which requires from Eq.~(\ref{m-sigma})
\be
\frac{\lambda}{\beta(1-\F)} < \gamma^2 \ll \lambda \,.
\ee

\begin{figure}
\centering
\includegraphics{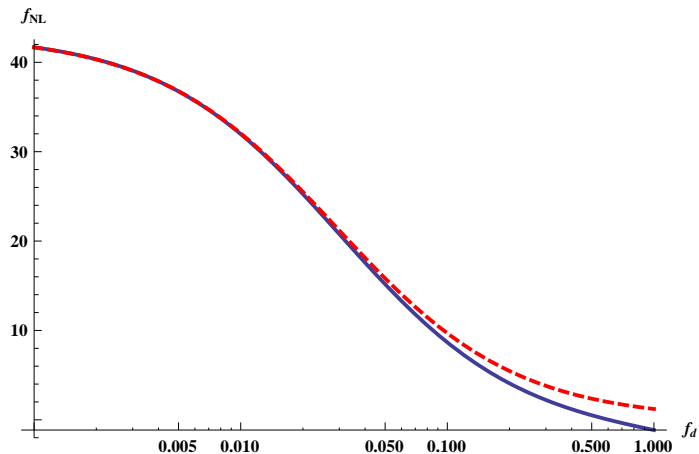}
\caption{The non-linearity parameter, $\fNL$ defined in Eq.~(\ref{fNL-def}), as a function of the curvaton density at decay, $f_d$, for a heavy curvaton after the end of inflation, with fixed curvaton density, $f_e=0.001$ and assuming $w_\sigma=1$. The exact result (\ref{zeta-final}) is shown by the solid blue line, while the approximate form (\ref{fNL-approx}) is shown by the dashed red line.}
\label{Heavy1}
\vspace{0.2 cm}
\end{figure}

\begin{figure}
\centering
\includegraphics{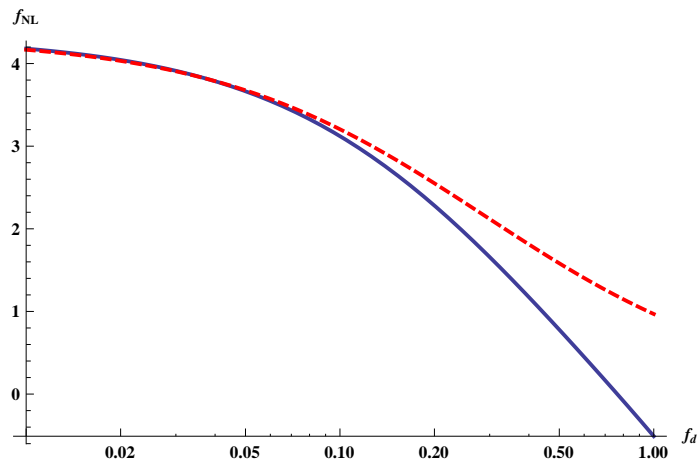}
\caption{The non-linearity parameter, $\fNL$, as a function of the curvaton density at decay, $f_d$, as in Figure~\ref{Heavy1} but for $f_e=0.01$.}
\label{Heavy2}
\vspace{0.5 cm}
\end{figure}

The curvaton first begins to oscillate at the end of inflation, when $\zeta=\zeta_e$, and from Eqs.~(\ref{def-zetagamma}) and~(\ref{def-zetasigma}) we have at this time
\ba
\zeta_\sigma - \zeta_\gamma
 &=& \left[ \frac{1}{3} \ln \left( 1+ \frac{\delta \rho_\sigma}{\rho_\sigma}
\right) - \frac{1}{4} \ln \left( 1+ \frac{\delta \rho_\gamma}{\rho_\gamma}
\right) \right]_{e} \,, \nonumber \\
   &=& \frac{2}{3(1-f_e)} \frac{\delta\sigma}{\sigma} - \left[ \frac{3(1-f_e)^2-f_e(3+5f_e)}{9(1-f_e)^2} \right] \left(\frac{\delta\sigma}{\sigma}\right)^2  +\ldots
\label{zeta-sigma-end}
 \,,
\ea
where, by analogy with Eq.~(\ref{f-def}), we identify
\ba
 \label{fe-def}
f_e \equiv \left( \frac{3\Omega_\sigma}{4-\Omega_\sigma} \right)_e = \frac{3\F(1-\F)}{2-\F(1-\F)} \,.
\ea
Combining Eqs.~(\ref{zeta-gamma}), (\ref{curvaton-zeta}) and (\ref{zeta-sigma-end}) we obtain the primordial density perturbation due to inflaton perturbations and due to fluctuations in the $\sigma$ field affecting the end of inflation and/or the curvaton effect due to late-decay
\ba
 \label{zeta-final}
 \zeta = \zeta_e + \frac{2}{3} \left( \frac{f_d-f_e}{1-f_e} \right) \frac{\delta\sigma}{\sigma}
 + \left( \frac{2f_d(1-f_d)(3+f_d)-3f_d(1-f_e)^2-f_e(1-f_d)(3+5f_e)}{9(1-f_e)^2} \right) \left( \frac{\delta\sigma}{\sigma} \right)^2 + \ldots
 \,.
\ea
where $\zeta_e$ is given by Eq.~(\ref{final-zeta-e}). Note that, since $f_d$ is an increasing function of time, we must have $f_d\geq f_e$.
In figures~\ref{Heavy1} and~\ref{Heavy2} we illustrate the corresponding non-Gaussianity parameter (\ref{fNL-def}) as a function of $f_d$ for two fixed values of $f_e$.

In the limit that the curvaton decays immediately after the end of inflation, $f_d\to f_e$, we recover $\zeta\to\zeta_e$, i.e., the density perturbation is exactly that given in Eq.~(\ref{final-zeta-e}) due to the inflaton perturbations plus the effect of the inhomogeneous end of inflation. On the other hand, for $f_e\ll f_d\leq1$ we have
\ba
 \zeta &\simeq& \zeta_e + \frac{2f_d}{3} \frac{\delta\sigma}{\sigma}
 + \left( \frac{f_d(3-4f_d-2f_d^2)}{9} \right) \left( \frac{\delta\sigma}{\sigma} \right)^2 + \ldots
 \,,
\nonumber\\
  &\simeq& \frac{3}{\alpha}\frac{\delta\phi_*}{\phi_*} + \left( \frac{2f_d}{3} + \frac{2f_e}{\alpha} \right) \frac{\delta\sigma}{\sigma}
 + \left( \frac{f_d(3-4f_d-2f_d^2)}{9} + \frac{f_e}{\alpha} \right) \left( \frac{\delta\sigma}{\sigma} \right)^2 + \ldots
 \label{zeta-final-limit}
 \,.
\ea
which is simply the sum of the usual curvaton perturbation plus the perturbations due to the inflaton and Lyth effect at the end of inflation. Note that for $f_d\gg f_e$ we require both $\Gamma\ll H_e$ and $f_e\ll1$, and hence from Eqs.~(\ref{fL}) and~(\ref{fe-def}) we have $\F\simeq 2f_e/3$ in the above expression.

For $f_d\gg f_e/\alpha$, the perturbation induced by the inhomogeneous end of inflation becomes negligible and we have
\be
\zeta \simeq \frac{3}{\alpha}\frac{\delta\phi_*}{\phi_*}
 + \frac{2f_d}{3} \frac{\delta\sigma}{\sigma}
 + \frac{f_d(3-4f_d-2f_d^2)}{9} \left( \frac{\delta\sigma}{\sigma} \right)^2 + \ldots
 \,.
 \label{zeta-heavy-simple}
 \ee
Note that for $f_e\ll1$ we have $f_d/f_e\propto a_d/a_e \propto (H_e/\Gamma)^{1/2}$ and thus for $f_d\gg f_e/\alpha$ we require $f_e\ll\alpha$ and $\Gamma\ll\alpha^2H_e$. If $f_d\geq\alpha$ or $\Gamma\geq\alpha^2H_e$ then the effect of the inhomogeneous end of inflation cannot be neglected with respect to the effect of the curvaton decay.

On the other hand, assuming only that the curvaton is very sub-dominant when it decays, $f_e\leq f_d \ll 1$, we obtain directly from Eqs.~(\ref{final-zeta-e}) and~(\ref{zeta-final}) that
\ba
\zeta &\simeq&
 \frac{3}{\alpha}\frac{\delta\phi_*}{\phi_*} + \left( \frac{2(f_d-f_e)}{3} + \frac{2f_e}{\alpha} \right) \frac{\delta\sigma}{\sigma}
 + \left( \frac{f_d-f_e}{3} + \frac{f_e}{\alpha} \right) \left( \frac{\delta\sigma}{\sigma} \right)^2 \,.
 \label{zeta-simple}
  \ea
We see that if the curvaton density at the end of inflation is non-negligible, then it can effect the primordial density perturbation not only through the inhomogeneous end of inflation, proportional to $f_e/\alpha$, but also by modifying the transfer parameter for the curvaton decay, from $f_d$ to $f_d-f_e$.

\subsection{Light curvaton after the end of inflation}

Alternatively, the curvaton effective mass may be smaller than the Hubble scale at the end of inflation, $m_\sigma<H_*$, which requires from Eq.~(\ref{m-sigma})
\be
\gamma^2 < \frac{\lambda}{\beta(1-\F)} \ll \lambda \,.
\ee
In this case the curvaton remain overdamped until the Hubble rate during the radiation era after inflation drops below the effective mass of the curvaton. For simplicity we assume that the curvaton remains fixed, $\sigma=\sigma_*$, until we have $H=m_\sigma$ at which point the field begins to oscillate and evolve like matter.

The radiation perturbation, $\zeta_\gamma$, after inflation is given by Eq.~(\ref{zeta-gamma}) and this remains constant until the curvaton decays.
On large scales, the uniform expansion hypersurface $H=m_\sigma$ corresponds to a uniform-density hypersurface, $\delta\rho_\gamma=-\delta\rho_\sigma$ and hence from Eqs.~(\ref{def-zetagamma}) and~(\ref{def-zetasigma}) we have
\ba
 \label{zeta-sg}
\zeta_\sigma - \zeta_\gamma
 &=& \left[ \frac{1}{3} \ln \left( 1+ \frac{\delta \rho_\sigma}{\rho_\sigma}
\right) - \frac{1}{4} \ln \left( 1+ \frac{\delta \rho_\gamma}{\rho_\gamma}
\right) \right]_{H= m_\sigma}\\
&=& \frac{2}{3 (1- f_o) } \frac{\delta \sigma}{\sigma}
 - \left[ \frac{3(1-f_o)^2-f_o(3+ 5 f_o)}{9 (1- f_o)^2} \right]
  \left(\frac{\delta \sigma}{\sigma}\right)^2 + \ldots \,,
\ea
where
\ba
f_o \equiv \left( \frac{3 \Omega_{\sigma}}{4-\Omega_{\sigma} } \right)_{H= m_\sigma} \,,
\ea
and we have $f_d\geq f_o\geq f_e$. After the curvaton starts oscillating it behaves like a pressureless fluid and the entropy perturbation, (\ref{zeta-sg}), remains constant on large scales until the curvaton decays into radiation. In the limit of a heavy curvaton that begins oscillating at the end of inflation, and hence $f_o=f_e$, we recover the previous result (\ref{zeta-sigma-end}).

\begin{figure}
\centering
\includegraphics{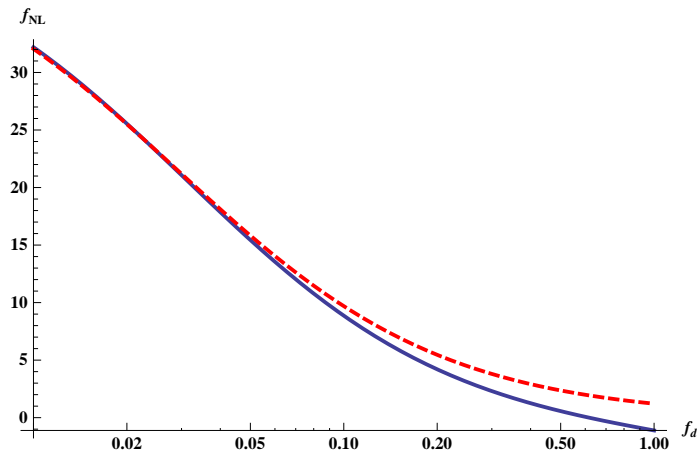}
\caption{The non-linearity parameter, $\fNL$ defined in Eq.~(\ref{fNL-def}), as a function of the curvaton density at decay, $f_d$, for a light curvaton after the end of inflation, with fixed curvaton density immediately after the end of inflation, $f_e=0.001$, and when the curvaton begins to oscillate, $f_o=0.01$, assuming $w_\sigma=1$. The exact result (\ref{zeta-final-light}) is shown by the solid blue line, while the approximate form (\ref{fNL-approx}) is shown by the dashed red line.}
\label{Light1}
\vspace{0.2 cm}
\end{figure}

\begin{figure}
\centering
\includegraphics{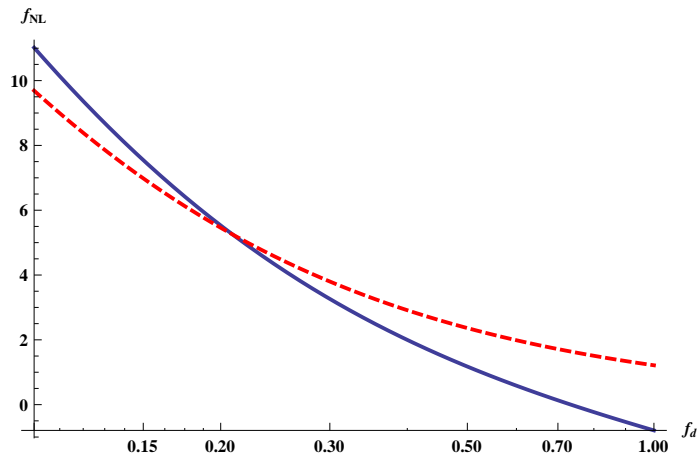}
\caption{The non-linearity parameter, $\fNL$, as a function of the curvaton density at decay, $f_d$, as in Figure~\ref{Light1} but for $f_e=0.001$ and $f_o=0.1$.}
\label{Light2}
\vspace{0.5 cm}
\end{figure}

Substituting Eqs.~(\ref{zeta-gamma}) and~(\ref{zeta-sg}) into Eq.~(\ref{curvaton-zeta}) gives the resulting primordial density perturbation after the curvaton decays
\ba
\zeta &=& \zeta_e
 + \frac{2}{3} \left[ \frac{f_d}{1- f_o} - \frac{f_e}{1- f_e}
\right] \frac{\delta \sigma}{\sigma}
 \nonumber\\
&&
 \quad
 + \left[ \frac{f_d \left( 2 (1- f_d) (3 +f_d) -3(1-f_o)^2 +f_o(3+5f_o) \right)}{ 9 (1- f_o)^2 } - \frac{f_e (3+ 5 f_e)}{ 9 (1- f_e)^2} +
\right] \left(\frac{\delta \sigma}{\sigma}\right)^2 + \ldots
 \,.
 \label{zeta-final-light}
\ea
Again one can check that in the limit $f_o=f_e$ we recover the previous result (\ref{zeta-final}).
In figures~\ref{Light1} and~\ref{Light2} we illustrate the corresponding non-Gaussianity parameter (\ref{fNL-def}) as a function of $f_d$ for fixed values of $f_e$ and two difference values of $f_o$.

If the curvaton mass is much smaller than the Hubble scale at the end of inflation ($m\ll H_e$) and the curvaton is very sub-dominant at the end of inflation, we have $f_e\ll f_o\leq f_d\leq 1$. In this case we can simplify Eq.~(\ref{zeta-final-light}) to give
\be
\zeta \simeq \zeta_e + \frac{2f_d}{3(1-f_o)} \frac{\delta \sigma}{\sigma}
 + \left[ \frac{f_d \left( 2 (1- f_d) (3 +f_d) -3(1-f_o)^2 +f_o(3+5f_o) \right)}{ 9 (1- f_o)^2 } \right]
  \left(\frac{\delta \sigma}{\sigma}\right)^2
\,.
  \ee
If in addition the curvaton is still very sub-dominant when it starts oscillating and the curvaton decay is slow relative to its mass ($\Gamma\ll m_\sigma$) so that $f_e\ll f_o\ll f_d\leq1$, then we recover the same result as we obtain for a heavy curvaton at the end of inflation when $f_e\ll f_d\leq1$, Eq.~(\ref{zeta-final-limit}).

On the other hand, assuming only that the curvaton is very sub-dominant when it decays, $f_e\leq f_o\leq f_d \ll 1$, we can recover from Eq.~(\ref{zeta-final-light}) the same simple expression (\ref{zeta-simple}). This expression is the same regardless of whether the curvaton is heavy or light immediately after inflation ends.

\section{Observational predictions}

We have shown how a curvaton field can contribute to the primordial density perturbation both through its effect on the end of inflation, in a hybrid inflation model, and through its final decay into radiation. The relative contribution of $\sigma$ to the power spectrum (\ref{power0}) can be written as
\ba
\label{def-w}
w_\sigma = \frac{N_{,\sigma}^2}{N_{,\phi}^2+N_{,\sigma}^2} \,.
\ea
Note that $N_{,\phi}=-H/\dot\phi\simeq-(2\epsilon)^{-1/2}M_P^{-1}$, hence the power spectrum is dominated by inflaton perturbations if $\epsilon\ll(2N_{,\sigma}^2M_P^2)^{-1}$ and dominated by perturbations in the $\sigma$-field if $\epsilon\gg(2N_{,\sigma}^2M_P^2)^{-1}$.

The spectral index is given in terms of slow-roll parameters as \cite{Wands:2002bn}
\be
n_\zeta - 1 \simeq -2\epsilon +(1-w_\sigma)(2\eta_\phi-4\epsilon)\, ,
\ee
where $\eta_\phi = M_P^2 V_{\phi \phi}/V$.
In the limit that the inflaton perturbations dominate the primordial power spectrum, $w_\sigma\ll1$, we recover the familiar result $n_\zeta\simeq-6\epsilon+2\eta_\phi$ and hybrid inflation models of the form given in Eq.~(\ref{potential}) produce a blue tilted spectrum, $n_\zeta>1$, in the vacuum-dominated limit~\cite{Copeland:1994vg}.

If $\sigma$-field perturbations dominate the primordial power spectrum, $w_\sigma\approx1$ due to either the inhomogeneous end of inflation or curvaton decay, then we have $n_\zeta-1\simeq-2\epsilon$ since the curvaton is massless during inflation. This is compatible with observational evidence favoring a red spectrum, $n_s<1$, and from Eq.~(\ref{epsilon}) we see that for $\alpha\approx0.5$ and $\phi\approx M_P$ we obtain $n_\zeta\approx0.97$.

The primordial non-Gaussianity (\ref{fNL-def}) can then be written as
\ba
\label{fNL-sigma}
\frac{6}{5} \fNL = w_\sigma^2 \frac{N_{,\sigma\sigma}}{N_{,\sigma}^2} \, .
\ea
If the $\sigma$-field perturbations have a red spectrum, and the inflaton perturbations have a blue spectrum then the relative contribution of the two different field perturbations to the primordial density perturbation will be scale-dependent. This can yield a scale-dependence of the non-linearity parameter (\ref{fNL-sigma}) \cite{Byrnes:2009pe,Byrnes:2010ft, Fonseca:2011aa}
\be
 n_{\fNL} \simeq 4(1-w_\sigma)(2\epsilon-\eta_\phi) \simeq -4\epsilon + 2(1-n_\zeta) \,.
\ee
If the inflaton perturbations dominate the primordial power spectra, $w_\sigma \ll 1$, then the non-linearity parameter, $\fNL$, is suppressed. On the other hand if the $\sigma$-field perturbations dominate, $w_\sigma \simeq 1$, then $\fNL$ could be large, but it also becomes scale-independent. Self-interactions of the curvaton field during inflation, which we have assumed to be negligible, could lead to a scale-dependence of the non-linearity parameter, without affecting the scalar tilt of the power spectra \cite{Byrnes:2010ft}.

In the following we will focus on the case $w_\sigma\approx1$ and $\epsilon\simeq0.02$, corresponding to $n_\zeta\simeq0.96$ and $n_{\fNL}\ll|n_\zeta-1|$.

\subsection{End of inflation}

Let us assume first of all that the curvaton decays soon after the end of inflation. This corresponds to a heavy curvaton at the end of inflation, $m_\sigma>H$, and a rapid decay, $f_d\simeq f_e$. Equation (\ref{zeta-final}) then reduces to $\zeta=\zeta_e$, i.e., the primordial density perturbation is given solely by the inhomogeneous end of inflation (\ref{final-zeta-e}).

The relative contribution of the $\sigma$ field to the primordial power spectrum (\ref{def-w}) is then
\be
 w_\sigma = \frac{\F^2\phi_*^2}{(1-\F)^2\sigma_*^2+\F^2\phi_*^2} \,.
 \ee
Thus we have $w_\sigma\approx1$ in the limit
\be
(1-\F)^2 \sigma_*^2 \ll \F^2 \phi_*^2 \,.
\ee
The $\sigma$ field perturbations will therefore dominate the primordial power spectrum so long as the $\sigma$ field VEV is much smaller than that of the inflaton field during inflation.

The non-linearity parameter (\ref{fNL-sigma}) is given by
\be
 \frac35 \fNL = w_\sigma^2 \frac{\alpha}{6}\frac{1+\F}{\F} \,.
\ee
In the limit $w_\sigma\approx1$ and $\F\ll1$ this can be re-written as
\be
 \label{fNL-end}
  \fNL \approx \frac{5\alpha}{9\Omega_{\sigma, e}} \,.
  \ee
This agrees with the original result of Lyth \cite{Lyth:2005qk}, but here we have expressed it in terms of the fractional density of the field immediately after inflation, in analogy with the usual result for a curvaton field.
Just as in the conventional curvaton scenario, we see that the non-Gaussianity becomes large in the limit that the curvaton density is small at the end of inflation.

If the $\sigma$-field decays rapidly at the end of inflation then the curvature perturbation is conserved on large scales throughout the radiation dominated era. However if the curvaton is sufficiently long-lived then it may further affect the primordial perturbation through the usual curvaton mechanism.

\subsection{Curvaton decay}


If we include the effect of the inhomogeneous decay of the curvaton sometime after the end of inflation then we may again get significant non-Gaussianity when the curvaton density is small, yet its contribution to the primordial power spectrum is significant. Assuming that the curvaton density remains sub-dominant throughout, $\Omega_{\sigma, d}$, we have from Eq.~(\ref{zeta-simple})
\be
 \label{fNL-approx}
 \fNL = w_\sigma^2 \frac{5}{4[f_d-f_e+(3f_e/\alpha)]} \,,
 \ee
where
\be
 w_\sigma = \frac{4\alpha^2[f_d-f_e+(3f_e/\alpha)]^2\phi^2}{81\sigma^2+4\alpha^2[f_d-f_e+(3f_e/\alpha)]^2\phi^2}
 = \frac{8\epsilon[f_d-f_e+(3f_e/\alpha)]^2\mPl^2}{9\sigma^2+8\epsilon[f_d-f_e+(3f_e/\alpha)]^2\mPl^2}
  \,.
 \ee
This approximate form for the non-Gaussianity is shown in figures~\ref{Heavy1} and~\ref{Heavy2} alongside the exact result, Eq.~(\ref{zeta-final}), for a heavy curvaton field after the end of inflation, and in figures~\ref{Light1} and~\ref{Light2} alongside the exact result, Eq.~(\ref{zeta-final-light}), for a light curvaton field after the end of inflation.

The existence of primordial perturbations due to the inhomogeneous end of inflation then places an upper bound on the non-Gaussianity possible in this case:
\be
\label{fNL-approx2}
 \fNL \leq \frac{5\alpha}{12f_e} \,.
\ee


Assuming the initial curvaton density is sufficiently small, $f_e\ll\alpha$, and the decay rate is slow enough, $\Gamma/H_e\ll \alpha^{2}$, then the effect of the curvaton field at the end of inflation can be neglected and the expression for the curvaton perturbation reduces to (\ref{zeta-heavy-simple}). The corresponding non-linearity parameter is
\be
 \frac35 \fNL = w_\sigma^2 \left( \frac{3-4f_d-2f_d^2}{4f_d} \right) \,.
 \ee
In the limit $f_d\ll1$ and $w_\sigma\simeq1$ this reduces to the famous result $\fNL\simeq 5/4f_d$. In terms of the fractional energy density when the curvaton decays,  Eq. (\ref{f-def}), this gives $\fNL\simeq 5/3\Omega_{\sigma, d}$.

\section{Discussion}

In this paper we have studied a model of hybrid inflation within which the primordial density perturbation can be produced either through an inhomogeneous end of inflation or through a curvaton scenario when the field decays, or from a combination of the two. The perturbations originate from vacuum fluctuations during inflation in a light isocurvature field, $\sigma$. This field is coupled to the waterfall field, $\chi$, whose tachyonic instability triggers the sudden end of hybrid inflation. Hence fluctuations, $\delta\sigma$, can affect the point at which the instability is triggered. Because the field $\sigma$ acquires a mass at the phase transition, it will begin oscillating about its minimum when the mass is larger that the Hubble scale after inflation. If the oscillations are sufficiently long-lived then the energy density in the $\sigma$ field may become significant and inhomogeneities in the energy density of the $\sigma$ field are transferred to the primordial radiation density when the field decays, as in the curvaton scenario.
\footnote{Our model is very different from the recently proposed ``hybrid curvaton'' model of Dimopoulos et al \cite{Dimopoulos:2012nj} where the curvaton field has a waterfall-type potential, so that oscillations of the curvaton field are triggered by a tachyonic instability.}

We are able to recover the standard results for primordial perturbations from an inhomogeneous end of inflation when the $\sigma$ field decays instantaneously at the end of inflation. We also recover familiar curvaton results in the limit where the field is sufficiently long-lived and curvature perturbations produced at the end of inflation are negligible. In both cases we express the resulting non-linearity parameter, $\fNL$, in terms of the fractional density of the curvaton field, $\Omega_\sigma$, either at the end of inflation or when the curvaton decays. We find that for large non-Gaussianity we have
\be
\fNL \approx
\left\{
\begin{array}{ll}
w_\sigma^2 ({5\alpha}/{9\Omega_{\sigma,e}}) & {\rm for}\ \Omega_{\sigma,e} \gg \alpha\Omega_{\sigma,d} \\
w_\sigma^2 ({5}/{3\Omega_{\sigma,d}}) & {\rm for}\ \Omega_{\sigma,d} \gg \Omega_{\sigma,e}/\alpha
\end{array}
\right.
 \,.
\ee
Any contribution to the overall primordial power spectrum from inflaton fluctuations, $w_\sigma<1$, tends to suppress the primordial non-Gaussianity.

Primordial perturbations from an inhomogeneous end of inflation are enhanced by the slow-roll parameter $\alpha$. Nonetheless curvaton-type perturbations tend to dominate over the end-of-inflation effect when the decay is slow, $\Gamma\ll \alpha^2 H_e$, unless the $\sigma$-density is already significant immediately after the end of inflation, $\Omega_{\sigma,e}>\alpha$.
More generally we find that perturbations due to the inhomogeneous end of inflation place an upper bound on the non-Gaussianity
\be
\fNL \leq \frac{5\alpha}{9\Omega_{\sigma,e}} \,.
\ee

We have used sudden-end-of-inflation and sudden-decay approximations to match the curvature and density perturbations across the end-of-inflation hypersurface and the curvaton-decay hypersurface. We expect this instantaneous matching to be a good approximation on length-scales much larger than the corresponding Hubble length. This long-wavelength approximation is usually an excellent approximation for scales relevant for the large-scale structure of our Universe \cite{Deruelle:1995kd,Malik:2006pm,Sasaki:2006kq}. On the other hand we know that the tachyonic instability in the waterfall field at the end of inflation leads to a very inhomogeneous fragmentation of the inflaton and waterfall field due to tachyonic preheating \cite{Felder:2000hj, Felder:2001kt}. If a curvaton field, $\sigma$, is weakly coupled to the waterfall field so that its mass remains less than the Hubble scale (a light curvaton immediately after inflation) then it remains overdamped and we would not expect the coherent curvaton field to be significantly disrupted by preheating in the inflaton and waterfall fields. On the other hand if the curvaton is sufficiently strongly coupled so that its mass becomes larger than the Hubble scale immediately after inflation, then its dynamics may be considerably more complicated, and possibly highly non-linear. Previous studies \cite{Enqvist:2008be,Chambers:2009ki} have considered resonant decay of the curvaton, but here we are considering possible parametric production of the curvaton itself. This has no effect on large scale density perturbations if all the fields rapidly thermalize, but short wavelength curvaton modes can alter the predictions in the curvaton scenario~\cite{Sasaki:2006kq}. A full study of the effect on the curvaton production and decay after tachyonic preheating in this intermediate regime requires a careful numerical treatment which goes well beyond the present work.

\section*{Acknowledgement}
We would like to thank A.~A.~Abolhasani,  H.~Bazrafshan, X.~Chen,  E.~Lim and  G.~Tasinato
for useful discussions. HF would like to thank the ICG, Portsmouth, for hospitality while this work was initiated and later finalized. MHN is in part supported by Yukawa Institute for Theoretical Physics (YITP), Kyoto University, under the Exchange
Program for Young Researchers of YITP.
DW is supported by STFC grant ST/H002774/1.



\end{document}